\documentclass{aastex}
\usepackage{spr-astr-addons}

\RequirePackage{color}

\begin{document}

\title{The influence of cosmic-rays on the magnetorotational instability}
%\slugcomment{Not to appear in Nonlearned J., 45.}
%% Running heads
\shorttitle{magnetorotational instability}
\shortauthors{Khajenabi}

\author{Fazeleh Khajenabi\altaffilmark{1}}
\affil{Department of Physics, Golestan University, Basij Square, Gorgan, Iran\\f.khajenabi@gu.ac.ir}

\begin{abstract}
We present a linear perturbation analysis of the magnetorotational instability in the presence of the cosmic rays. Dynamical effects of the cosmic rays are considered by a fluid description and the diffusion of  cosmic rays is only along the magnetic field lines. We show an enhancement in the growth rate of the unstable mode because of the existence of cosmic rays. But as the diffusion of cosmic rays increases, we see that the growth rate decreases. Thus, cosmic rays have a destabilizing  role in the magnetorotational instability of the accretion discs.
\end{abstract}

\keywords{galaxies: active - black hole: physics - accretion discs}

%\section*{}
%\label{sec:intro}

\section{Introduction}
Understanding the true nature of accretion processes in astrophysics has always been an attractive research topic over the last  three decades. Accretion discs are observed in many astrophysical systems from  new born stars to compact objects or even very large discs at the center of the galaxies. In spite of the diversity of the accreting systems, existence of a possible mechanism of the angular momentum transport is a common feature in all these accretion systems. Extensive efforts to understand mechanisms of the angular momentum transport in the accretion discs have lead to a better understanding of such systems, though there are many theoretical and observational uncertainties.

It has been proposed that the  magnetorotational instability (MRI) is the main driving mechanism of turbulence in the accretion discs \citep{Balbus1991}. Extensive subsequent works have clarified and extended our understanding of the role of MRI in various astrophysical systems from protoplanetary discs \citep[e.g.,][]{sano99} to the protostellar discs or even quasar discs. Over recent years a multi layer model for the protoplanetary discs is proposed in which the surface layers are magnetically active due to the ionization of the CRs, while the central layers are magnetically inactive because of inability of CRs to penetrate down to the central parts \citep[e.g.,][]{gammie}. Thus, MRI can act as a driving mechanism of the turbulence and the accretion at the surface layer in a protoplanetary disc. However, possible {\it dynamical} effects of CRs on MRI have not been studied to our knowledge.

Cosmic Rays are very energetic particles but their energy density is in equipartition with energy densities of thermal gas and turbulence \citep[e.g.,][]{fer2001}. CRs can act as a source of heating and increase the level of the ionization in the interstellar medium \citep[e.g.,][]{field69}. An enhanced flux of CRs has important consequences for star formation near to the Galactic center \citep{yusef2007}.

However, interaction of CRs with a plasma is not restricted just to a possible enhancement of the level of ionization as have been studied extensively over recent decades. For example, dynamical effect of CRs has a vital role in analysis of Parker instability for the structure formation in the Galaxy at  large scale \citep[e.g.,][]{parker66,mous74,hanasz2000,kuwabara2004,kuwabara2006}. It is also found that CRs have a stabilizing effect on the thermal instability \citep{kuz83, wagner2005, shadmehri2009}.

The problem of the  diffusion of CRs and its role in MRI has not been studied in detail. Considering CRs as a separate  fluid and their diffusion along the magnetic field lines, we study MRI in the presence of CRs via a linear perturbation analysis. Our basic equations and the assumptions are presented in the next section. Final dispersion relation will be analyzed in sections 3 and 4.

\section{General Formulation}
\label{sec:GF}
In this study, CRs are protons, electrons and nuclei. But we neglect the electrons because of their little contribution to the total pressure.
There are three different approaches to study the dynamics of CRs. In the particle-particle approach, the plasma and CRs  are considered as particles that may interact with each other via complicated processes. In a simpler approach, known as fluid-particle, the plasma is treated as a fluid, though CRs are still described as particles. The simplest approach is the fluid-fluid approach in which CRs and the thermal gas are described by different interacting fluids. The hydrodynamic  approach can not provide the spectrum of CRs, however, it is a good approximation for analyzing  dynamics of a plasma with CRs \citep[e.g.,][]{drury81,drury83}. Thus, we adopt the hydrodynamic approach to study the effects of CRs on the unstable modes of MRI.

We also follow the same steps as in, except  that CRs are considered as a fluid, and diffusion is considered only along magnetic field lines. For simplicity, we neglect the ionization and the heating by the CRs, since their effects in the absence of the dynamical role of CRs are well understood.

The basic equations are
\begin{equation}\label{eq:cont}
\frac{d\rho}{dt}=-\rho \nabla \cdot {\bf v},
\end{equation}
\begin{equation}\label{eq:motion}
\rho \frac{d{\bf v}}{dt}=\frac{1}{4\pi} (\nabla \times {\bf B}) \times {\bf B} - \nabla (p+p_{\rm cr}),
\end{equation}
\begin{equation}
\frac{\partial {\bf B}}{\partial t}=\nabla \times ({\bf v} \times {\bf B}),
\end{equation}
\begin{equation}
\nabla \cdot {\bf B}=0,
\end{equation}
\begin{equation}\label{eq:energyg}
\rho \frac{de}{dt}=-p \nabla \cdot {\bf v} + \Lambda ,
\end{equation}
and
\begin{equation}\label{eq:CRS}
\frac{1}{\gamma_{\rm cr}-1}\frac{dp_{\rm cr}}{dt}-\frac{\gamma_{\rm cr}}{\gamma_{\rm cr}-1} \frac{p_{\rm cr}}{\rho} \frac{d\rho}{dt}+\nabla \cdot {\bf \Gamma} =0,
\end{equation}
where $d/dt = \partial / \partial t + {\bf v} \cdot \nabla$ is the Lagrangian time derivative. Here, $\rho$, $p$, $p_{\rm cr}$, ${\bf v}$ and ${\bf B}$ are
the density, gas pressure, cosmic ray pressure, velocity and the magnetic field, respectively. Also, the net cooling function is denoted by $\Lambda$ in the
energy equation (\ref{eq:energyg}). However, we will neglect the net cooling in this paper, i.e. $\Lambda =0$. Moreover, we have
\begin{equation}
{\bf \Gamma}=-\kappa_{\parallel} {\bf b} ({\bf b} \cdot \nabla p_{\rm cr}),
\end{equation}
is the diffusive flux of cosmic-ray energy and $\kappa_{\parallel}$ is diffusion coefficient along magnetic field lines. All the variables have their usual meaning. Also, $\bf b$ is a unit
 vector along the magnetic field lines, i.e. ${\bf b}= {\bf B}/B$. The adiabatic indices of the thermal gas and cosmic rays are denoted by $\gamma_{\rm g}$ and $\gamma_{\rm cr}$, respectively.
Finally, we can write equation of state as
\begin{equation}\label{eq:STATE}
p=\frac{R}{\mu} \rho T,
\end{equation}
where $R$ is the gas constant and $\mu$ represents the molecular weight.

\section{Linear Perturbations}
Assuming that the system is axisymmetric, we write basic equations in the cylindrical coordinates $(R, \Phi, z)$. The equilibrium magnetic field is assumed to be constant in space and only the toroidal and the vertical components of the magnetic field are considered, i.e. ${\bf B_{0}}=B_{\rm z} {\bf e}_{\rm z} + B_{\Phi} {\bf e}_{\Phi}$. The equilibrium disc is rotating with Keplerian angular velocity, i.e. $\Omega = \sqrt{GM/R^{3}}$ where  $M$ is the mass of the central object. Moreover, the initial density and the gas  and the cosmic ray pressures are considered to be constant.  Now, we can  perturb the equations of the form
$\delta X = \delta X_{0} \exp i(k_{\rm R}R + k_{\rm z}z - \omega t)$.   Thus, the linearized equations become
\begin{equation}
-i \omega \frac{\delta \rho}{\rho} + i k_{\rm R} \delta v_{\rm R} + i k_{\rm z} \delta v_{\rm z} =0,
\end{equation}
\begin{displaymath}
-i\omega \delta v_{\rm R} - 2 \Omega \delta v_{\Phi} + \frac{i k_{\rm R}}{\rho} \delta p + \frac{i k_{\rm R}}{\rho} \delta p_{\rm cr}
\end{displaymath}

\begin{equation}
-\frac{i k_{\rm z}}{4\pi\rho} B_{\rm z} \delta B_{\rm R} + \frac{i k_{\rm R}}{4\pi\rho} (B_{\Phi} \delta B_{\Phi} + B_{\rm z} \delta B_{\rm z}) =0,
\end{equation}
\begin{equation}
-i \omega \delta v_{\Phi} + \frac{\kappa^{2}}{2\Omega} \delta v_{\rm R} - \frac{i k_{\rm z}}{4\pi\rho} B_{\rm z} \delta B_{\Phi} =0,
\end{equation}
\begin{equation}
-i \omega \delta v_{\rm z} + \frac{ik_{\rm z}}{\rho} \delta p + \frac{ik_{\rm z}}{\rho} \delta p_{\rm cr} + \frac{ik_{\rm z}}{4\pi \rho} B_{\theta} \delta B_{\theta}=0,
\end{equation}
\begin{equation}
-i \omega \delta B_{\rm R} - i k_{\rm z} B_{\rm z} \delta v_{\rm R} =0,
\end{equation}
\begin{displaymath}
-i \omega \delta B_{\Phi} - i k_{\rm z} B_{\rm z} \delta v_{\Phi} - \frac{d \Omega}{d \ln R} \delta B_{\rm R}+i k_{\rm R} B_{\Phi} \delta v_{\rm R}
\end{displaymath}
\begin{equation}
 + i k_{\rm z} B_{\Phi} \delta v_{\rm z} =0,
\end{equation}
\begin{equation}
-i \omega \delta B_{\rm z} + i k_{\rm R} B_{\rm z} \delta v_{\rm R} =0,
\end{equation}
\begin{equation}
-i \omega \frac{\delta p}{p} + i \omega \gamma_{\rm g} \frac{\delta\rho}{\rho} =0,
\end{equation}
\begin{equation}
\left [ \frac{-i \omega}{\gamma_{\rm cr} -1} + \kappa_{\parallel} ({\bf b}.{\bf k})^{2} \right ] \delta p_{\rm cr} + \frac{\gamma_{\rm cr}}{\gamma_{\rm cr} -1} \frac{p_{\rm cr}}{\rho} i \omega \delta \rho =0,
\end{equation}
where $\kappa^{2}$ is the square of the epicyclic frequency,
\begin{equation}
\kappa^{2}=\frac{2\Omega}{R}\frac{d}{dR}(R^{2}\Omega).
\end{equation}

\begin{figure*}%[tb]
\includegraphics[scale=0.8]{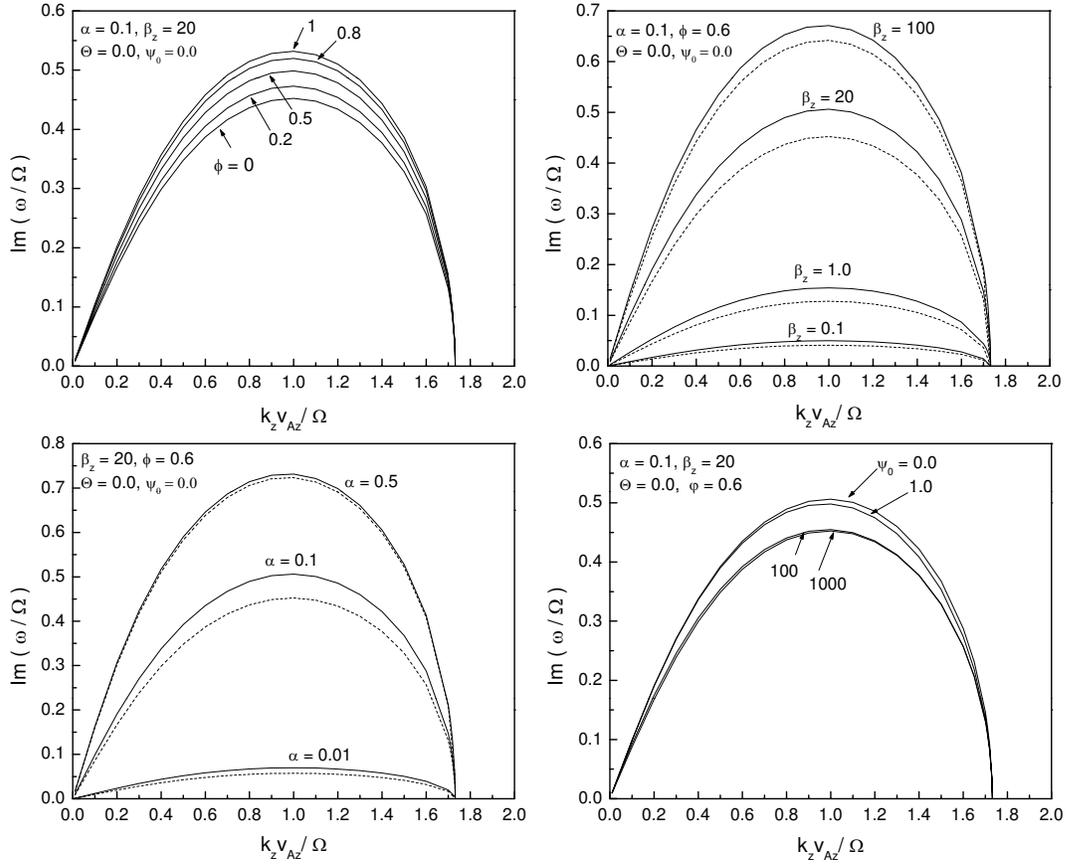}
\caption{Imaginary part of the growth rate as a function of the vertical wave number}
\label{fig:f1}
\end{figure*}

Introducing a new dimensionless variable as $X=\omega /(i \Omega)$, our final dispersion equation becomes
\begin{displaymath}
P_{7} X^{7} + P_{6} X^{6} + P_{5} X^{5} + P_{4} X^{4} + P_{3} X^{3}
\end{displaymath}
\begin{equation}
P_{2} X^{2} + P_{1} X + P_{0} =0,\label{eq:disper}
\end{equation}
where the coefficients are complicated function of the input parameters. We denote the angle between the vector ${\bf k}$ and the component $k_{z}$ by $\Theta$. Also, a nondimensional wavenumber
is defined as $Z=k_{\rm z} V_{\rm Az}/\Omega$. Thus, the coefficients become
%(see appendix A for the complete forms of the coefficients).
%
\begin{equation}
{P_{0}} = 2 \,{\gamma _{\rm g}}\, \Lambda \,\,\psi \,q\, \alpha ^{2} \Omega^{2} {c_{0}}^{3} Z^{5} \,({\gamma _{\mathit{\rm cr}}} - 1),
\end{equation}
\begin{equation}
{P_{1}} = 2\,\Lambda\,\,
 \alpha ^{2} \Omega^{2} \,\, \,{c_{0}}^{2}\, Z^{4} ({\gamma _{\rm g}} + {\gamma _{\mathit{\rm cr}}}\,\phi  ) \mathrm{cos} \Theta ,
\end{equation}
\begin{displaymath}
P_{2}=2 [ (1+c_{0}^{2}\alpha^{2})(\kappa^{2}/\Omega^{2})\cos^{2}\Theta + \alpha^{2}\Lambda + (1+
\end{displaymath}
\begin{equation}
2c_{0}^{2}\alpha^2)Z^{2}] ({\gamma _{\mathit{cr}}} - 1)\, \Omega^{2} Z^{3}\,{\gamma _{g}}\,{c_{0}}\,\psi \,q
\end{equation}
\begin{displaymath}
P_{3}=-Z^{2} \Omega^{2} \cos^{2}\Theta \{ \gamma_{\rm g} c_{0} \psi q Z \alpha (\gamma_{\rm cr} -1)(4+2\lambda
\end{displaymath}
\begin{displaymath}
 - \kappa^{2}/\Omega^{2}) \sin \Theta - 2\cos^{2}\Theta [c_{0}^{2}\alpha^{2} (\gamma_{\rm g} + \gamma_{\rm cr} \varphi) \kappa^{2}/\Omega^{2}
\end{displaymath}
\begin{displaymath}
+ \gamma_{\rm g} (2\lambda \alpha^{2} + \kappa^{2}/\Omega^{2})] - 2Z^{2} [2c_{0}^{2} \alpha^{2} (\gamma_{\rm g} + \gamma_{\rm cr} \varphi)
\end{displaymath}
\begin{equation}
+ \gamma_{\rm g} (1+\alpha^{2})] \}
\end{equation}
\begin{displaymath}
P_{4}= \gamma_{\rm g} Z \Omega^{2} [ \alpha (\kappa^{2}/\Omega^{2}-2\lambda - 4) Z \cos^{2}\Theta \sin\Theta + 2c_{0} \times
\end{displaymath}
\begin{displaymath}
 \psi_{0} q \alpha^{2} (\gamma_{\rm cr}-1) (Z^2 - \kappa^{2}/\Omega^{2}) \cos^{2}\Theta + 2c_{0} \psi_{0} q (\gamma_{\rm cr} -1)\times
\end{displaymath}
\begin{equation}
(1+\alpha^{2}+c_{0}^{2}\alpha^{2}) Z^{2}]
\end{equation}
\begin{displaymath}
P_{5}= 2 \Omega^{2} [ \gamma_{\rm g} \alpha^{2} (Z^{2}+ \kappa^{2}/\Omega^{2}) \cos^{2}\Theta + \gamma_{\rm g} Z^{2} (1+\alpha^{2}) +
\end{displaymath}
\begin{equation}
Z^{2} c_{0}^{2}\alpha^{2} (\gamma_{\rm g} + \gamma_{\rm cr} \varphi) ] \cos\Theta
\end{equation}

\begin{equation}
{P_{6}} =2\,{\gamma _{\rm g}}\,{c_{0}}\,\psi_{0} \,q\,\,\alpha ^{2} \Omega ^{2}\,\,\, Z ({\gamma _{\mathit{cr}}} - 1) \cos^{2}\Theta
\end{equation}

\begin{equation}
{P_{7}} =2\,{\gamma _{\rm g}}\,\,\alpha^{2} \Omega ^{2}\, \cos^{3}\Theta .
\end{equation}
where $\Lambda = 2\lambda \cos^{2}\Theta + Z^2$. We verified that the above dispersion equation reduces to the equation for a case without CRs (e.g., see Eq. (17) of \citet{sano99}). In the above relations, we have $\lambda=-d(\ln\Omega)/d(\ln R)$, $q=[{\bf b}.({\bf k}/k)]^{2}$. Also, the parameter $\varphi$ represents the ratio of the pressure of CRs to the gas pressure, i.e. $\varphi = p_{\rm cr} / p$. The initial direction of the magnetic field is denoted by $\alpha = B_{\rm z} / B_{\Phi}$.
The plasma beta parameter is defined by the poloidal field, $\beta_{\rm z} = 2 c_{\rm s}^{2} / \gamma_{\rm g} v_{\rm Az}^{2}$ where $c_{\rm s}$ is the sound speed and $v_{\rm Az}$ is the Alfven velocity, i.e. $v_{\rm Az}=B_{\rm z}/\sqrt{4\pi\rho}$. Thus, we obtain $c_{\rm s} = c_{0} v_{\rm Az}$, where $c_{0}=\sqrt{(\gamma_{\rm g}/2)\beta_{\rm z}}$. Also, we have $\psi = \kappa_{\parallel} k / c_{\rm s} = \psi_{0}Z/\cos\Theta $, where the nondimensional diffusion  coefficient $\psi_{0}$ is
\begin{equation}
\psi_{0}=c_{0}^{2} \left ( \frac{\tau_{c}}{\tau_{\rm D}} \right ) \left ( \frac{\tau_{\rm c}}{\tau_{\rm d}} \right ),
\end{equation}
where the diffusion time scale $\tau_{\rm D}$, the dynamical time scale $\tau_{d}$ and the sound crossing time scale $\tau_{\rm c}$ are defined as
\begin{equation}
\tau_{\rm D} = \frac{R^{2}}{\kappa_{\parallel}}, \tau_{\rm d}=\frac{1}{\Omega}, \tau_{\rm c}= \frac{R}{c_{\rm s}}.
\end{equation}

\section{analysis}
Equation (\ref{eq:disper}) describes magnetorotational instability with CRs. Considering complexity of the coefficients of our dispersion equation, it is very unlikely to obtain the roots in analytical closed forms. However, we can solve the equation numerically for a disc with the Keplerian angular velocity. Since our goal is to analyze the unstable perturbations, we restrict our study to the roots with positive imaginary part, i.e. ${\rm Im} (\omega) >0$. If we neglect terms corresponding to the CRs, i.e. $\phi=\psi_{0}=0$,  our algebraic dispersion relation (\ref{eq:disper}) reduces to the classical dispersion equation \citep[e.g.,][]{sano99}. Among our input parameters, the effects of CRs are described via two input parameters $\phi$ and $\psi_{0}$. One can vary these parameters to study how the unstable modes are modified due to the existence of CRs.

Typical influence of CR on the magnetorotational instability are shown in all subsequent plots. Figure \ref{fig:f1} shows nondimesnional growth rate of the unstable mode, $\omega/\Omega$, versus nondimensional wavenumber $Z$. Each curve is labeled by its corresponding parameter. We found the maximum growth rate occurs for the perturbations with $k_{\rm R}=0$ \citep[see also,][]{sano99}.  The top left-hand plot of Figure \ref{fig:f1} shows  behavior of the unstable perturbation when the ratio of the CRs pressure to the gas pressure $\varphi$ changes from zero to one with $\psi_{0}=0$, $\Theta=0$, $\beta_{\rm z}=20$ and $\alpha=0.1$. Here, diffusion of CRs is neglected. Obviously, the case with $\varphi=0$ corresponds to the unstable mode without CRs. As the ratio $\varphi$ increases the system becomes more unstable because of the enhancement of the growth rate. In other words, existence of the CRs destabilizes the disc.

The top right-hand plot of Figure \ref{fig:f1} shows the typical dependence of the growth rate on the parameter $\beta_{z}$ which is ratio of the gas pressure to the magnetic pressure. The input parameters are $\psi_{0}=0$, $\Theta=0$, $\varphi=0.6$ and $\alpha=0.1$ and $\beta_{z}$ varies  from low value $0.1$ to high value $100$. Corresponding to  each case represented by the solid lines, there is a dashed curve which is for the same case but without CRs. The bottom left-hand plot of Figure \ref{fig:f1} shows the case with CRs (solid line) and without CRs (dashed line) for different values of  $\alpha$ which is the ratio of the $z$ component of magnetic filed to the $\varphi$ component of the magnetic filed.  The parameter $\alpha$ changes from low value $0.01$ to high value $0.5$ and the other input parameters are  $\psi_{0}=0$, $\Theta=0$, $\varphi=0.6$ and $\beta_{z}=20$. In all previous plots, diffusion of CRs along the magnetic field lines is neglected. Now, the bottom right-hand plot of Figure \ref{fig:f1} shows how the growth rate of the unstable perturbation is modified when the diffusion of CRs is not negligible. Here, all the input parameters are fixed as $\alpha=0.1$, $\Theta=0$, $\varphi=0.6$ and $\beta_{z}=20$ but the dimensionless diffusion parameter  $\psi_{0}$ changes from $0$ to $1000$. Diffusion of CRs has a stabilizing effect according to this plot. However, when diffusion is high we can hardly recognize any changes and the plots overlap.

\section{conclusion}
Our simple approach shows that CRs destabilizes MRI unstable modes. It implies that the generated turbulence because of the MRI would be amplified in the presence of the CRs. However, it seems that growth rate only slightly modifies in the presence of CRs in the linear regime. In our model, gas and CRs coupling is via cosmic ray pressure term in the equation of motion. In other words, CRs provides an extra pressure and we know the pressure plays minor roles in MRI so long as the magnetic pressure is smaller than the gas and the CRs pressures. But when the diffusion of CRs along the magnetic field lines increases, CRs pressure decreases and the system tends to a case without CRs.  Thus, it would be interesting to study MRI with CRs at the nonlinear regime by doing numerical simulations.  There are regions with high flux of CRs such as near to the Galactic center or ultra luminous infrared galaxies \citep{papa2010,papa2011}.
Under such circumstances, we think, evolution of the accretion discs located at such regions are significantly affected by the dynamical effects of CRs.

\section*{Acknowledgments}

I am grateful to the anonymous referee whose detailed and careful comments helped to improve the quality of this paper. I would also like to thank Peter Duffy and Luke Drury for helpful comments.

%\section*{Acknowledgments}
%
%We are grateful to the anonymous referee whose comments helped to improve the quality of this paper.

\bibliographystyle{spr-mp-nameyear-cnd}
%\bibliography{myref}
\bibliography{referenceKHr}

\end{document}